\documentclass[12pt]{iopart}

\usepackage{graphicx}
\usepackage{epstopdf}
\newcommand{\del}[0]{\partial}

\newtheorem{definition}{Definition}

\newcommand{\bra}[1]{\langle {#1} \vert}
\newcommand{\ket}[1]{\vert {#1} \rangle}
\newcommand{\text}{\mathrm}

\newcommand{\LC}{\mathcal{L}}
\newcommand{\MC}{\mathcal{M}}
\newcommand{\OC}{\mathcal{O}}

\newcommand{\RC}{\mathcal{R}}

\newcommand{\WC}{\mathcal{W}}

\newcommand{\dya}[1]{\ket{#1}\!\bra{#1}}

\begin{document}

\title[Large gradients via correlation in random parameterized quantum circuits]{Large gradients via correlation in random parameterized quantum circuits}

\author{Tyler Volkoff}

\address{Theoretical Division, Los Alamos National Laboratory, Los Alamos, NM, USA.}

\author{Patrick J. Coles}

\address{Theoretical Division, Los Alamos National Laboratory, Los Alamos, NM, USA.}
\vspace{10pt}

\begin{abstract}
Scaling of variational quantum algorithms to large problem sizes requires efficient optimization of random parameterized quantum circuits. For such circuits with uncorrelated parameters, the presence of exponentially vanishing gradients in cost function landscapes is an obstacle to  optimization by gradient descent methods. In this work, we prove that reducing the dimensionality of the parameter space by utilizing circuit modules containing spatially or temporally correlated gate layers can allow one to circumvent the vanishing gradient phenomenon. Examples are drawn from random separable circuits and asymptotically optimal variational versions of Grover's algorithm based on the quantum alternating operator ansatz (QAOA). In the latter scenario, our bounds on cost function variation imply a transition between vanishing gradients and efficient trainability as the number of layers is increased toward $\OC(2^{n/2})$, the optimal oracle complexity of quantum unstructured search.
\end{abstract}

%
%
%
%
%

\section{Introduction}\label{sc:intro}

Variational quantum algorithms are a class of quantum algorithms especially suited to near-term applications including eigenvalue estimation~\cite{mcclean2016,peruzzo2014VQE,jones2019variational,nakanishi2019subspace,larose2019variational,cerezo2020variational,arrasmith2019variational}, quantum compiling~\cite{qaqc,sharma2020noise}, linear systems~\cite{bravo2019variational,xu2019variational,linlin}, and quantum dynamics~\cite{li2017efficient,yuan2018theory,heya2019subspace,cirstoiu2019variational,otten2019noise}. They consist of a random parameterized quantum circuit (RPQC) module coupled via quantum measurement to a classical module for cost function optimization. The RPQC component of a variational quantum algorithm is constructed by contracting layers of correlated or uncorrelated parameterized quantum gates in a sequence that depends on the application at hand. Correlated gate layers in RPQCs, or, more generally, in quantum neural networks, can be motivated by the task of generating target states with desired symmetries or coherence properties, whereas uncorrelated gates are often utilized to simulate random unitary operations or scramble quantum information.  Examples of quantum algorithm modules involving application of correlated gate layers to a compound quantum register~\cite{watrousbook} include coherence generation by a tensor product of Hadamard gates in quantum phase estimation~\cite{nc} and quantum algorithms for linear equations~\cite{linsys}, and sequential application of identical two-qubit unitary operators for generating translationally invariant quantum states~\cite{PhysRevLett.95.110503,cirac}.

The classical module of a variational quantum algorithm involves a classical decision problem which determines that an estimate of the cost function either satisfies a condition for optimality or requires further optimization according to an update rule for the RPQC. When the update rule involves a gradient descent-based optimization step, the presence of ``barren plateau landscapes'' (BPL)~\cite{mcclean2018barren} in cost functions of generic variational quantum algorithms presents a challenge for efficient optimization of RPQCs ~\cite{kubler2019adaptive,arrasmith2020operator,harrow2019,sweke2019}. One strategy for circumventing BPL in variational quantum algorithms utilizing hardware-efficient RPQCs (e.g., bricklayer circuits \cite{PhysRevB.100.134306} with depth scaling as $\log n$) consists of using local operators to define the cost function \cite{cfdbp}. For short-depth variational quantum algorithms that admit a faithful cost function defined by a sum of local observables, this strategy increases the efficiency of the RPQC update and hence the trainability of the algorithm. However, for quantum algorithms that require polynomial \cite{PhysRevLett.97.050401} or exponential circuit depth \cite{PhysRevLett.79.325}, or require the use of a  cost function defined by a non-local observable (e.g., quantum communication protocols that rely on collective measurements for their efficiency \cite{gotchuang}), variational versions require new approaches for avoiding BPL.

In this work, we show that it is possible to avoid BPL in certain variational quantum algorithms and algorithm modules by using RPQC architectures containing correlated parameters, even when the cost function is defined by a global operator such as a projection onto a pure state of the full register. Our first examples include variational quantum compiling with spatially correlated, single qubit gate layers (Section \ref{sec:qub}) and Haar random $m$-qubit gates (Section \ref{sec:hrand}). The rest of our analysis is concerned with circuits inspired by the quantum alternating operator ansatz (QAOA) \cite{farhi1}. We show that BPL are avoided in a quantum approximate optimization algorithm for a simple MaxCut problem when either local cost functions are used or when global cost functions are combined with correlated ansatz parameters and large circuit depth (Section \ref{sec:lcf} and  \ref{sec:rod}). 

Finally, our main result consists of a proof that QAOA-inspired variational versions of Grover's algorithm \cite{rieffel,biamonte} exhibit a transition from BPL at low circuit depths to trainability at circuit depths that coincide with a high algorithm success rate (Section \ref{sec:uns}). In precise terms, if the number of oracle applications scales as $2^{cn - \log_{2}n}$, with $0<c<1/3$, then the variational quantum search exhibits BPL. However, if the number of oracle applications scales as $2^{cn - \log_{2}n}$, with $c>1/2$, then the variational quantum search does not exhibit BPL. We conclude that trainability of these variational versions of Grover's algorithm requires a circuit depth that coincides (up to a logarithmic correction) with the optimal oracle complexity of quantum unstructured search. Both interlayer and spatial correlation of parameters is crucial to this result; removing the constraint of interlayer parameter correlation implies BPL for circuit depths coinciding with the optimal oracle complexity.

\section{Background}

The mathematical setting for studies of BPL in variational quantum algorithms is a parameterized set of $n$-qubit quantum states $\lbrace \ket{\psi(\theta)}: \theta \in \Omega \subset \mathbf{R}^{m} \rbrace \subset \mathbf{C}^{2^{n}}$, where $\Omega$ is a compact set of parameters equipped with a probability density $p:\Omega \rightarrow [0,1]$. Performance of the algorithm is quantified by a cost function random variable 
\begin{equation}X(\theta)=\langle \psi(\theta)\vert O \vert \psi(\theta)\rangle,
\label{eqn:cf1}
\end{equation} where $O$ is a bounded, self-adjoint operator. In this work, we examine the following property:

\begin{definition} (Barren plateau landscape)
The cost function $X(\theta)$ exhibits a BPL with respect to $\theta_{j}$ if it is continuously differentiable on a compact subset $\mathcal{A} \subset \Omega$ of the parameter space and if for every $\epsilon >0$, there exists $0<b<1$ such that $P_{\mathcal{A}}(\vert{\del X \over \del \theta_{j}}\vert \ge \epsilon) \in \OC(b^{n})$, where $P_{\mathcal{A}}$ is the probability measure on $\mathcal{A}$ induced from $p$. \end{definition}

We often choose $\mathcal{A}$ such that $X(\theta)$ has global minima in $\mathcal{A}$. In this case, presence of the BPL for $\mathcal{A}$ precludes efficient trainability of the variational quantum algorithm by gradient descent. Further, one may be interested in training a submodule of an RPQC, in which case the subset $\mathcal{A}$ defines the parameter space of the submodule.

\begin{figure*}[t]
    \centering
    \includegraphics[scale=0.5]{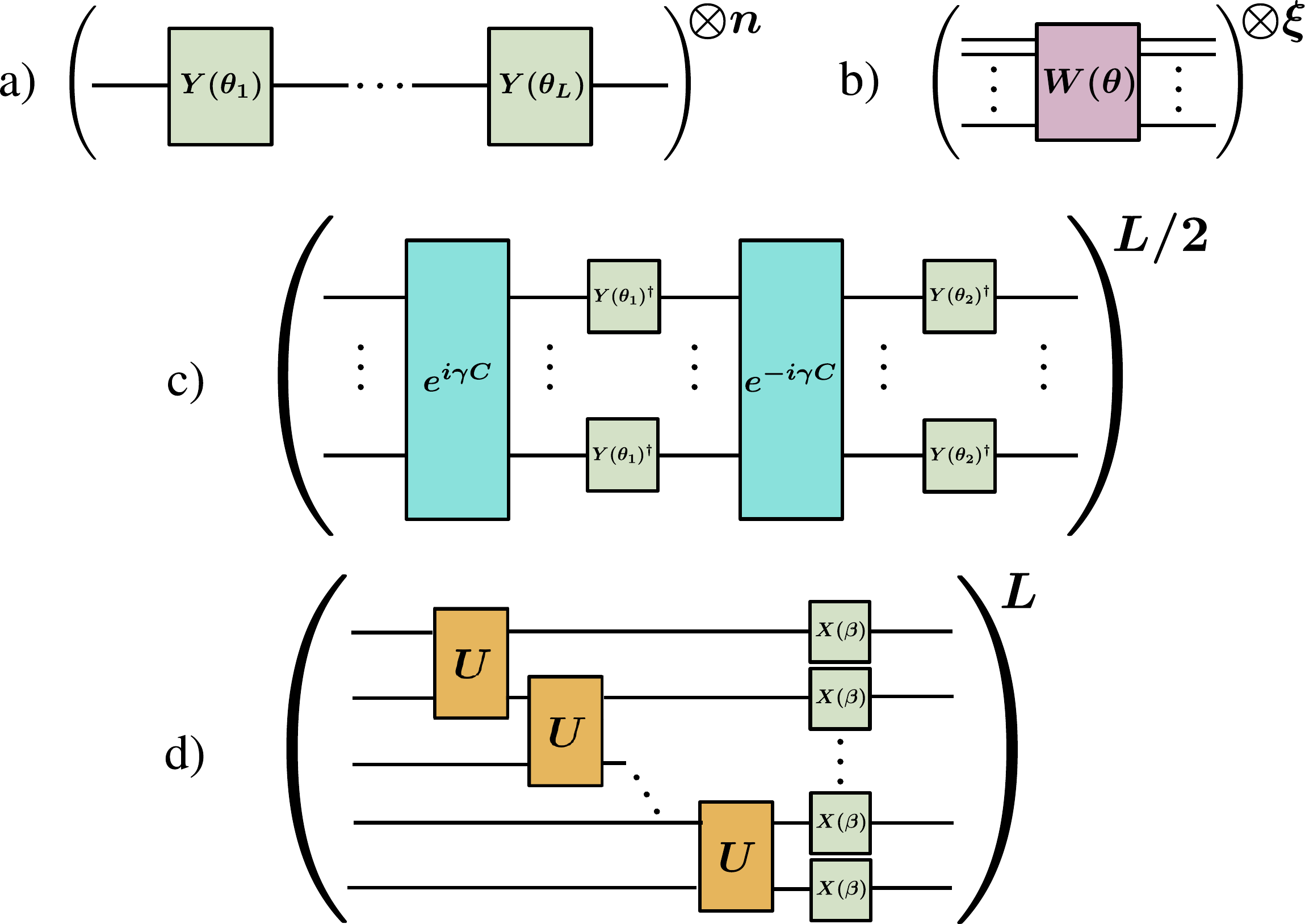}
    \caption{RPQCs for circuit modules with correlated parameters. a) $\MC(\theta)$ in Section \ref{sec:qub} with $Y(\theta):=e^{-i{\theta \over 2}\sigma_{y}}$. b)  $\WC(\theta)$ in Section \ref{sec:hrand}. c) $\MC(\theta,\gamma)_{L}$ in the less efficient variational version of Grover's algorithm. Each layer $j\in \lbrace 1,\ldots ,L/2\rbrace$ is associated with parameters $(\theta_{2j-1},\theta_{2j},\gamma)$. d) $\MC(\beta,\gamma)_{L}$ in quantum alternating operator ansatz for the ring-of-disagrees problem analyzed in \ref{sec:rod} with correlated mixer layers and correlated driver layers. A local factor in the driver layer is given by $U=e^{i{\gamma \over 2}\sigma_{z}\otimes \sigma_{z}}$, and a local factor in the mixer layer is given by $X(\beta):= e^{-i\beta \sigma_{x}}$.  }
    \label{fig:first}
\end{figure*}

BPL are expected in randomly initialized quantum neural networks of sufficient depth \cite{mcclean2018barren,entbp}, but can also occur for RPQCs of low depth. If a given cost function does not exhibit BPL, one can consider that cost function to exhibit a large gradient (relative to the generic case). Because many quantum algorithms utilize layers having the same structure, one can search for variational versions of these algorithms by using RPQCs with spatially correlated or temporally correlated layers. For example, spatially correlated quantum circuits (e.g., permutation invariant, translation invariant, etc.)  appear in the transition functions that define 1-D quantum cellular automata \cite{PhysRevA.72.052301}, in algorithms for quantum state transfer \cite{PhysRevLett.97.090502}, and quantum convolutional neural networks \cite{choilukin}. 

\subsection{Circumventing the BPL problem}
In this work, we analyze variational quantum algorithm modules for which the strategy of spatially or temporally correlating parameters allows to circumvent BPL, even when the cost function is obtained by a global measurement of the variational state. This strategy can also avoid BPL for full variational quantum algorithms, as we show for the case of two variational versions of Grover's algorithm for quantum unstructured search.  In general, the strategy proposed in the present work of using spatially or temporally correlated RPQC to avoid BPL can be considered as an initialization scheme. A different initialization scheme proposed in Ref.~\cite{Grant2019initialization} circumvents generic BPL by compiling identity operator blocks in the RPQC, thereby delaying or avoiding sampling from a unitary 2-design. The initialization scheme in Ref.~\cite{verdon2019learning} involves temporal RPQC updates computed by an auxiliary classical neural network, and the question of trainability is relegated to analyses of gradients of the loss function of the classical neural network.  The ansatz initialization scheme in Ref.~\cite{fujii} assumes that the RPQC is a Clifford circuit for some parameter value, which implies that cost function derivatives can locally be computed efficiently.   When utilizing a given initialization scheme, the particular problem instances and circuit ansatz\"{e} should be taken into account, as has been emphasized in the case of variational quantum algorithms for linear systems of equations \cite{reben}. Other methods for increasing trainability of RPQCs include optimization of subsets of parameters \cite{adaptprun,layerwise}. We also note that certain quantum algorithm modules, e.g., quantum convolutional neural networks \cite{qcnnabs} or linear optical transformations of coherent states \cite{linoptabs}, have been shown to exhibit large gradients even in generic cases.

\section{\label{sec:qub}Large gradients in separable circuits}
\subsection{\label{sec:gcf}Global cost functions}

In the task of variational quantum estimation of the ground state energy of a $n$-qubit Hamiltonian $H$ with spectrum in $[0,1]$, one can define a real cost function random variable by 
\begin{equation}C_{\RC(\theta)}=\langle \psi(\theta) \vert H\vert \psi(\theta)\rangle \label{eqn:gc}\end{equation} where $\ket{\psi(\theta)}=\RC(\theta)\ket{0}^{\otimes n}$, $\RC:[-\pi,\pi)^{\times M}\rightarrow U(2^{n})$ is a RPQC that maps $M$ angles to the unitary group on $n$ qubits. We assume that there exists $\theta_{0}$ such that ground state of $H$ takes the form $\ket{\psi(\theta_{0})}$, since this assumption implies both the faithfulness of the cost function and the fact that the orbit of the RPQC on $\ket{0}^{\otimes n}$ contains the ground state. 

To illustrate the BPL phenomenon for the cost function (\ref{eqn:gc}), we first analyze the simple case when $H$ is taken to be the global projection $H=\mathbf{I}-\dya{0}^{\otimes n}$, which allows (\ref{eqn:gc}) to also be considered as the cost function for local variational quantum compiling of the identity operator~\cite{qaqc,sharma2020noise}. When the non-identity component of the operator $H$ in (\ref{eqn:gc}) is a sum of self-adjoint operators that act non-trivially on every qubit, we refer to $H$ as a global cost function. Consider a sequence of single qubit gates $V^{(j)}(\theta):=\prod_{i=1}^{L}e^{-i{\theta_{i}^{(j)}\over 2}\sigma_{y}}$, $j=1,\ldots,n$, that constitute an $n$-qubit RPQC $\LC(\theta)=\bigotimes_{j=1}^{n}V^{(j)}(\theta)$. We take $\lbrace\theta_{i}^{(j)}\rbrace$ to be a set of $L n$ independent, uniform random variables on $(-\pi , \pi]$.    In \cite{cfdbp} it was shown that, with respect to uniformly distributed $\theta$, the variance of $\del_{\nu}C_{\LC(\theta)}$ vanishes exponentially with $n$ (i.e., as $b^{n}$ with $0<b<1$), where $\del_{\nu}$ symbolizes the partial derivative with respect to any angle argument in $C_{\LC(\theta)}$. Since $E\left( \del_{\nu}C_{\LC(\theta)}\right)=0$, Chebyshev's inequality 
implies
 \begin{eqnarray}
P\left( \vert \del_{\nu}C_{\LC(\theta)} \vert \ge\epsilon \right)&\le& {E(\left( \del_{\nu}C_{\LC(\theta)} \right)^{2}) \over \epsilon^{2}} \nonumber \\
&=& {b^{n}\over \epsilon^{2}}.
\label{eqn:cheby}
\end{eqnarray} Inequality (\ref{eqn:cheby}) implies that the gradient is (almost everywhere) exponentially concentrated at 0, which is the defining feature of the BPL phenomenon.

\begin{figure}[t]
    \centering
    \includegraphics[width=.94\columnwidth]{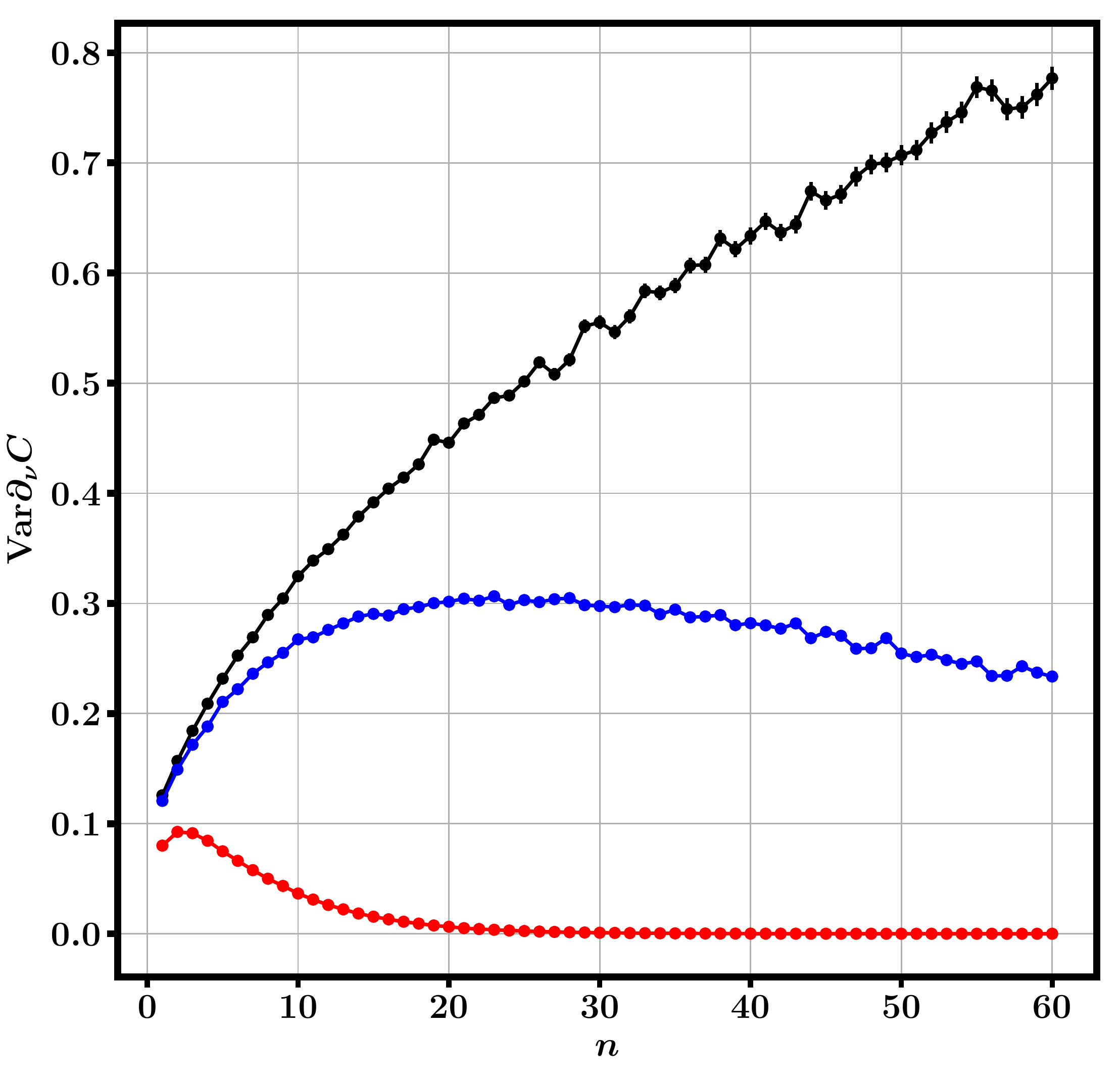}
    \caption{Monte Carlo integration of $\text{Var}\left( \del_{\nu}C_{\MC(\theta)} \right)$ with $L=4$ for cost function (\ref{eqn:glob}) (black) and cost function (\ref{eqn:delt})  with $\delta=0.01$ (blue) and $\delta = 0.10$ (red) for $n=1,\ldots , 60$ qubits. The asymptotic $\OC(\sqrt{n})$ behavior can be seen in the top trace, as predicted by (\ref{eqn:cvv}). The blue and red traces are asymptotically exponentially vanishing, indicating BPL for the cost function in (\ref{eqn:delt}). $5\times 10^{4}$ samples are used for each $n$.}
\label{fig:o}
\end{figure}

Alternatively, a simpler quantum circuit can be constructed by perfectly correlating the angles $\theta_{i}^{(j)}$ for all layers $j$, forming the permutation invariant circuit $\MC(\theta):= \left(\prod_{i=1}^{L}e^{-i{\theta_{i}\over 2}\sigma_{y}}\right)^{\otimes n}$ (see Fig.~\ref{fig:first}a). The circuit $\MC(\theta)$ can be considered as a $L$-dimensional submanifold of the $nL$-dimensional manifold that defines $\LC(\theta)$. Importantly, when the circuit input consists of a tensor product $\ket{\phi}^{\otimes n}$ with $\ket{\phi}$ chosen from the $xz$-plane of the Bloch sphere, the correlated RPQC $\MC(\theta)$ does not lose descriptive power compared to $\LC(\theta)$ for variational quantum compiling of the ground state $\ket{0}^{\otimes n}$ of Hamiltonian $H=\mathbf{I}-\ket{0}\bra{0}^{\otimes n}$, because the cost function still attains the value 0. For example, with $H=\mathbf{I}-\ket{0}\bra{0}^{\otimes n}$ and $\ket{\psi(\theta)}= \mathcal{M}(\theta)\ket{0}^{\otimes n}$, the cost function (\ref{eqn:gc}) is
\begin{equation}
C_{\MC(\theta)}=1-\cos^{2n}\left({\theta_{1}+\ldots +\theta_{L} \over 2}\right)
\label{eqn:glob}
\end{equation}
which attains the minimum value 0 at the $(L-1)$-dimensional critical submanifold defined by $\sum_{i=1}^{L}\theta_{i}=0$ mod  $2\pi$.  Because $E\left( \del_{\nu}C_{\MC(\theta)} \right)=0$, where $\del_{\nu}$ again symbolizes a partial derivative with respect to an argument of $C_{\MC(\theta)}$,
we find that (see \ref{sec:app1})
\begin{eqnarray}
\text{Var}\left( \del_{\nu}C_{\MC(\theta)} \right) &=& E(\left(\del_{\nu}C_{\MC(\theta)}\right)^{2}) \sim {n^{1\over 2}\over 4\sqrt{2\pi }}. 
\label{eqn:cvv}
\end{eqnarray}
Because (\ref{eqn:cvv}) is polynomially increasing with $n$, there is no BPL phenomenon for this variational quantum compiling task. However
the scaling in (\ref{eqn:cvv}) is valid only for pure input states. For instance, taking an input state $\rho^{\otimes n}$ with $\rho=\text{diag}(1-\delta,\delta)$, where $\delta<1/2$, one finds that 
\begin{equation}
C_{\MC(\theta)} =1-\left( \delta - (1-2\delta)\cos^{2}\left({\theta_{1}+\ldots +\theta_{L} \over 2}\right)  \right)^{n}.
\label{eqn:delt}
\end{equation}
Figure~\ref{fig:o} shows the results of Monte Carlo integration over $[-\pi,\pi)^{\times L}$ with $L=4$ of $\text{Var}\left( \del_{\nu}C_{\MC(\theta)} \right)$ for cost functions (\ref{eqn:glob}) and (\ref{eqn:delt}) with $\delta=0.01$ and $\delta = 0.10$ for $n=1,\ldots , 60$ qubits. For $\delta = 0.01$, which corresponds to input states close to the pure state manifold, the variance of the gradient of (\ref{eqn:delt}) does not increase asymptotically, but instead exhibits a crossover. The BPL phenomenon is clearly seen in the curve in Fig.~\ref{fig:o} corresponding to $\delta=0.10$ in  (\ref{eqn:delt}). The data show that even when the RPQC parameters are correlated, trainability of a variational quantum algorithm can depend sensitively on the purity of the input register. In fact, a recent analysis of cost functions for a large class of noisy RPQCs (e.g., noisy QAOA) indicate the appearance of BPL for noisy circuit depths scaling at least linearly in $n$ with a noise-dependent coefficient \cite{noiseinduced}. The result holds even when parameters are correlated in the RPQC.

Although the circuit $\MC(\theta)$ is permutation invariant, it is not necessary that the input state or global projection $H$ have any symmetry under a subgroup of the symmetric group in order for the BPL phenomenon to be absent for $C_{\MC(\theta)}$. For example, if $U$ is a unitary such that $[\MC(\theta),U]=0$, the variational quantum compiling cost function (\ref{eqn:gc}) is invariant under $H\mapsto UHU^{*}$ and $\ket{0}^{\otimes n}\mapsto U\ket{0}^{\otimes n}$.

The BPL phenomenon can also be avoided in spatially correlated, separable RPQC when the circuit layers do not commute. As an example, consider the variational quantum compiling task defined by the RPQC  \begin{equation}
\RC(\beta,\alpha)_{L}=\displaystyle \overleftarrow{\prod_{j=1,\ldots,L}}e^{-i{\beta_{j}\over \sqrt{n}}J_{x}}e^{-i{\alpha_{j}\over \sqrt{n}}J_{y}},
\label{eqn:rpqc1}
\end{equation}
where $\alpha_{j}$ and $\beta_{j}$ are uniformly distributed on $[-\pi ,\pi)$ for all $j$. We take a global cost function having the form of (\ref{eqn:cf1}) with $H=\dya{0}^{\otimes n}$.
The RPQC in (\ref{eqn:rpqc1}) generates an asymptotically normal quantum statistical model, i.e., the circuit generates normally distributed rotations from the uniform random angles $\beta$ and $\alpha$ \cite{PhysRevA.73.052108}. The cost function is asymptotically equal to
\begin{equation}
C_{\RC(\beta,\alpha)_{L}}\sim \cos^{2n}\left( {\sqrt{  \left(\sum_{j=1}^{L}\beta_{j} \right)^{2} +\left(\sum_{j=1}^{L}\alpha_{j}  \right)^{2}} \over 2\sqrt{n} }\right)
\label{eqn:asympcost}
\end{equation}
for large $n$. Equation~(\ref{eqn:asympcost}) is nonzero and independent of $n$ as $n\rightarrow \infty$, therefore it cannot exhibit the BPL phenomenon. Adding temporal correlations to the RPQC by taking, e.g., a subset of the $\alpha_{j}$ to be equal, only increases $\vert \del_{\alpha_{j}}C_{\RC(\beta,\alpha)_{L}}\vert$, thereby enhancing trainability.

\subsection{\label{sec:lcf}Local cost functions}

In contrast to a global cost function, a local cost function is defined by an $H$ in (\ref{eqn:gc}) which is a sum of operators that each act trivially on at least one qubit. Often, each operator in the sum that defines a local cost function has a small support, e.g., on $k\ll n$ qubits.
In \cite{cfdbp}, faithful local cost functions were shown to circumvent the BPL phenomenon for variational quantum compiling when the RPQC is in a class of hardware-efficient quantum circuits containing uncorrelated gates, including such circuits as $\LC(\theta)$. Local cost functions also allow to circumvent the BPL phenomenon in the case of the correlated RPQC $\MC(\theta)$. This can be seen by taking Hamiltonian $H=\dya{0}^{\otimes n}$, input state $\ket{0}^{\otimes n}$, and RPQC $\MC(\theta)$, and defining a faithful local cost function via $C^{(L)}_{\MC(\theta)}:=1-{1\over n}\sum_{j=1}^{n}\text{tr}\MC(\theta)\dya{0}^{\otimes n}\MC(\theta)^{\dagger}O_{j}$, where $O_{j}:= \dya{0}_{j} \otimes \mathbf{I}_{\overline{j}}$. Due to permutation invariance of the RPQC, it is clear that $C^{(L)}_{\MC(\theta)}$ is independent of $n$, and it follows that the variance of $\del_{\nu}C^{(L)}_{\MC(\theta)}$ is independent of $n$, which precludes BPL behavior. 

In a less trivial setting, we show in  \ref{sec:rod} that a local cost function for a simple quantum approximate optimization algorithm (namely MaxCut on regular, degree 2, connected graphs, i.e., the ring of disagrees~\cite{farhi1}) does not exhibit BPL. The RPQC of this variational algorithm consists of alternating applications of a translation invariant layer of correlated two-qubit gates and a layer of spatially correlated single qubit rotations, as shown in Fig.~\ref{fig:first}d. If a global cost function is used instead, e.g., by defining the cost function via a projection onto the subspace of maximum cut states, then BPL is avoided only if the RPQC has correlated parameters between layers of the same structure, and the RPQC has depth exponential in $n$. This suggests that when faithful local cost functions exist, they are preferable for training quantum approximate optimization algorithms with uncorrelated layers.

\section{\label{sec:hrand}Large gradients in $\xi$-separable circuits}

Whereas the RPQC considered in Section \ref{sec:qub} involved correlated single qubit gates, the RPQCs of principal interest in variational quantum algorithms contain layers with multi-qubit gates, e.g., quantum data bus ansatze for variational quantum state preparation \cite{kuzmin}, or layered hardware-efficient ansatze that appear in variational quantum algorithms for spectrum estimation \cite{cerezo2020variational}. For such applications, a global cost function can be defined as in (\ref{eqn:gc}), except the RPQC involves non-local unitary operations. In this section, we show that the BPL phenomenon is avoided in a generic correlated separable setting, namely when $n$ qubit registers are grouped into $\xi$ registers of $m$ qubits ($n=\xi m$  with $m,\xi \in \mathbf{N}$)  and the RPQC is a tensor product of Haar-distributed unitaries which is invariant under permutation of the $m$-qubit registers (see Fig.\ref{fig:first}b). 

We consider variational quantum estimation of the ground state energy of the global observable $H=\bigotimes_{j=1}^{\xi}O_{j}$, where $O_{j}$ are self-adjoint linear operators on $(\mathbf{C}^{2})^{\otimes m}$ with operator norm 1, and the input state is $\tau = \bigotimes_{j=1}^{\xi}\rho_{j}$. The global cost function is given by \begin{equation}
C_{\WC(\theta)} :=1-\text{tr}H\WC(\theta)\tau \WC(\theta)^{\dagger}
\end{equation}
where $\WC(\theta)$ is defined RPQC that has the translation-invariant form $\WC(\theta)= W(\theta)^{\otimes \xi}$, where $W(\theta)$ is a unitary on $m$-qubits. The structure of $\WC(\theta)$ is taken to have the same form as in Ref.~\cite{mcclean2018barren}, and is motivated by the design of a uniformly random unitary operation on an $m$-qubit register. Specifically,  $W(\theta)=\prod_{i=1}^{\zeta}e^{-i{\theta_{i}\over 2}\sigma_{f(i)}}G_{i}$, where $G_{i} \in U(2^{m})$ is an unparameterized unitary and  $f:\lbrace 1,\ldots ,\zeta \rbrace\rightarrow \lbrace 1,\ldots ,m\rbrace$ is a given function. The generator $\sigma_{f(i)}$ is a Pauli matrix acting on qubit register $f(i)$.   A uniform distribution on the $m$-qubit Clifford group constitutes a 2-design for $U(2^{m})$ and it has been shown that 2-designs on $m$ qubits can be implemented by using $\OC(m^{2})$ gates from a generating set of one- and two-qubit gates \cite{datahiding,dankert}. Local random quantum circuits on $m$ qubits consisting of $\OC(t^{10}m^{2})$ gates suffice to simulate approximate unitary $t$-designs \cite{brandaohoro}.

With the structure of the RPQC fixed, we find that 
\begin{eqnarray}
\del_{\nu}C_{\WC(\theta)}&=&{-i\over 2}\sum_{\ell=1}^{\xi}\left[ \left( \prod_{h\neq \ell}\text{tr}W_{B}\rho_{h}W_{B}^{\dagger}W_{A}^{\dagger}O_{h}W_{A} \right) \right. \nonumber \\  &{}& \left. \text{tr}W_{B}\rho_{\ell}W_{B}^{\dagger}[\sigma_{f(\nu)},W_{A}^{\dagger}O_{\ell}W_{A}] \vphantom{\prod_{h\neq \ell}}\right]
\label{eqn:del}
\end{eqnarray}
where $W_{A}=\prod_{i=\nu}^{\zeta}e^{-i{\theta_{i}\over 2}\sigma_{f(i)}}G_{i}$, $W_{B}=\prod_{i=1}^{\nu -1}e^{-i{\theta_{i}\over 2}\sigma_{f(i)}}G_{i}$, $W(\theta)=W_{A}W_{B}$, and $\del_{\nu}$ is a partial derivative with respect to any element of $\lbrace \theta_{i}\rbrace_{i=1}^{\zeta}$.

For the special case of input state $\tau=\rho^{\otimes \xi}$, and observable $H=O^{\otimes \xi}$, (\ref{eqn:del}) simplifies to
\begin{eqnarray}
\del_{\nu}C_{\WC(\theta)}&={-i\xi\over 2}\left( \text{tr}W_{B}\rho W_{B}^{\dagger}W_{A}^{\dagger}OW_{A} \right)^{\xi-1} \text{tr}W_{B}\rho W_{B}^{\dagger}[\sigma_{f(\nu)},W_{A}^{\dagger}OW_{A}].
\label{eqn:s2}
\end{eqnarray}
Due to the fact that for unitaries $W_{A}$ and $W_{B}$ the second trace factor in (\ref{eqn:s2}) has modulus between $0$ and $4$, we relabel the modulus of the second trace factor as $c$. The first trace factor determines the scaling of $\text{Var}\del_{\nu}C_{\WC(\theta)}$ with $\xi$. In particular, one finds that
\begin{equation}
c^{-2}\text{Var}\del_{\nu}C_{\WC(\theta)} = {\xi^{2}\over 4}\Bigg\vert E \left( \text{tr}W_{B}\rho W_{B}^{\dagger}W_{A}^{\dagger}OW_{A} \right)^{2(\xi-1)} \Bigg\vert
\label{eqn:pow}
\end{equation}
where $E:= E_{W_{A},W_{B}}$ is the expectation with respect to normalized Haar measure on $U(2^{m})$. To carry out the expectation over $W_{B}$, we use Egorychev's method to write the right hand side of (\ref{eqn:pow}) as a contour integral over a circle of radius $\epsilon$ centered at the origin
\begin{eqnarray}
{}&{}&c^{-2}E_{W_{B}}\left( (\del_{\nu}C_{\WC(\theta)})^{2}\right)  = {\xi^{2}\left( 2(\xi-1) \right)!\over 4}\nonumber \\
&\cdot& \Bigg\vert {1\over 2\pi i} \int_{\vert t\vert=\epsilon}dt \, t^{-(2\xi-1)}\int d\mu(W_{B})e^{t\text{tr}X_{A}W_{B}\rho W_{B}^{\dagger}} \Bigg\vert
\label{eqn:pow2}
\end{eqnarray}
where $X_{A}:=W_{A}^{\dagger}OW_{A}$.

The integral over Haar measure in (\ref{eqn:pow2}) is an example of a Harish-Chandra-Itzykson-Zuber integral and can be evaluated explicitly. Since the spectrum of $X_{A}$ is the same as the spectrum of $O$, the right hand side of (\ref{eqn:pow2}) is independent of $W_{A}$, and therefore the final result for $\text{Var}\del_{\nu}C_{\MC(\theta)}$ is
\begin{eqnarray}
{}&{}&c^{-2}\text{Var}\del_{\nu}C_{\WC(\theta)} = {\xi^{2}\left( 2(\xi-1) \right)!q\over 4\Delta(\lambda(\rho))\Delta(\lambda(O))}\nonumber \\
&{}&\cdot\Bigg\vert {1\over 2\pi i} \int_{\vert t\vert=\epsilon}dt \, t^{-(2\xi-1+{2^{m}\choose 2})}\det \left[ e^{t\lambda_{i}(O)\lambda_{j}(\rho)} \right]_{i,j} \Bigg\vert
\label{eqn:varsep}
\end{eqnarray}
where for any self-adjoint linear operator $A$ on $m$ qubits, $\lambda(A)$ is defined as the vector of eigenvalues of $A$ in ascending order, $\Delta(\lambda(X))=\prod_{1\le i <j \le 2^{m}}\lambda_{j}(A)-\lambda_{i}(A)$ is the Vandermonde determinant, and $q:=\prod_{j=1}^{2^{m}-1}j!$ is a constant.

The scaling of the right hand side of (\ref{eqn:varsep}) can be readity calculated in the simple case of $m=1$, $\rho=\text{diag}(1-\delta,\delta)$, where $0<\delta<1/2$, and $O=\dya{0}$. In this case, (\ref{eqn:varsep}) becomes (now with $n=\xi$, since $m=1$)
\begin{eqnarray}
c^{-2}\text{Var}\del_{\nu}C_{\WC(\theta)} &=& {\xi^{2}\left( 2(\xi-1) \right)!\over 4(1-2\delta)} \Big\vert {1\over 2\pi i} \int_{\vert t\vert=\epsilon}dt \, t^{-2\xi}e^{t(1-2\delta)}  \Big\vert \nonumber \\
&=&{\xi^{2}(1-2\delta)^{2\xi-2}\over 4(2\xi-1)}
\end{eqnarray}
where the residue theorem was used to evaluate the integral over the circular contour. Similar to the result of Section \ref{sec:qub}, one finds that the BPL phenomenon is avoided for pure input state ($\delta=0$), but is encountered for $\delta \neq 0$. To maintain a constant variance as $\xi \rightarrow \infty$, the input state $\rho$ can be taken with $\delta={\log \xi \over 4\xi -4}$, i.e., with purity scaling as
\begin{equation}
\text{tr}\rho^{2}\sim \left( 1-{\log \xi \over 4\xi -4 } \right)^{2}.
\end{equation}

\section{\label{sec:uns}Trainability of variational algorithms for unstructured search}

We now analyze BPL for variational quantum algorithms for unstructured search. A near-optimal version of Grover's algorithm has been proposed which utilizes a quantum circuit consisting of alternating applications of the Grover oracle $V=\mathbf{I}-2\dya{0}^{\otimes n}$ and a local rotation $U=e^{-{2\pi i\over n}J_{y}}$ \cite{rieffel}, where
\begin{equation}
J_{y}:={1\over 2}\sum_{j=1}^{n}\mathbf{I}_{1}\otimes \cdots \otimes \mathbf{I}_{j-1}\otimes \sigma_{y} \otimes \mathbf{I}_{j+1}\otimes \cdots \otimes \mathbf{I}_{n}.
\end{equation}
The alternating structure of the circuit is reminiscent of quantum algorithms based on the quantum alternating operator ansatz (QAOA) \cite{farhi1,hadfield2019quantum}. However, in QAOA, the unitary $V$ traditionally takes the form $V=\prod_{j=1}^{k}V_{j}$, where $\lbrace V_{j}\rbrace_{j}$ is a set of commuting unitaries. To formulate a variational version of the near-optimal Grover's algorithm in \cite{rieffel}, one may consider an RPQC of the alternating form
\begin{equation}
\RC(\alpha,\gamma)_{L}:=\displaystyle\overleftarrow{\prod_{k=1,\ldots,L}}e^{i\alpha_{k} J_{y}}e^{i\gamma_{k}\dya{0}^{\otimes n}}\,.
\label{eqn:ggrpqc}
\end{equation} 
Although the RPQC in (\ref{eqn:ggrpqc}) contains the optimal circuit in Ref.~\cite{rieffel}, it is not clear that it can be efficiently optimized. This is is especially true given that short depth sequences of alternating unitaries (e.g., one application of a unitary of the form $UV$ where $U$, $V$ are $2^{n}\times 2^{n}$ unitary matrices) generically require large depth sequences of alternating unitaries to be compiled using gradient descent optimization (e.g., unitaries of the form $(WT)^{d^{2}}$ where $W$, $T$ are random $2^{n}\times 2^{n}$ unitary matrices and $d$ is the Hilbert space dimension) \cite{lloydgrad}.

In this section we show that by taking parameters in $\RC(\alpha,\gamma)_{L}$ to be equal among layers with the same structure, and thereby reducing the dimension of the parameter space to two, the BPL phenomenon can be avoided in variational Grover's algorithm for sufficient circuit depth. Conversely, failing to correlate the parameters in this way necessarily leads to BPL during circuit optimization. These results suggest that the optimal submanifold of parameters in variational versions of Grover's algorithm also defines a trainable submanifold, i.e., a parameter space in which the optimum can be efficiently found by gradient descent methods. We note that although the original Grover's algorithm is optimal in the sense of having the minimal number of applications of the oracle operation \cite{PhysRevA.60.2746}, variational versions of Grover's algorithm are likely to be useful in development of quantum search algorithms with optimal total depth complexity \cite{korepin}.

\begin{figure*}[t]
    \centering
    \includegraphics[scale=0.5]{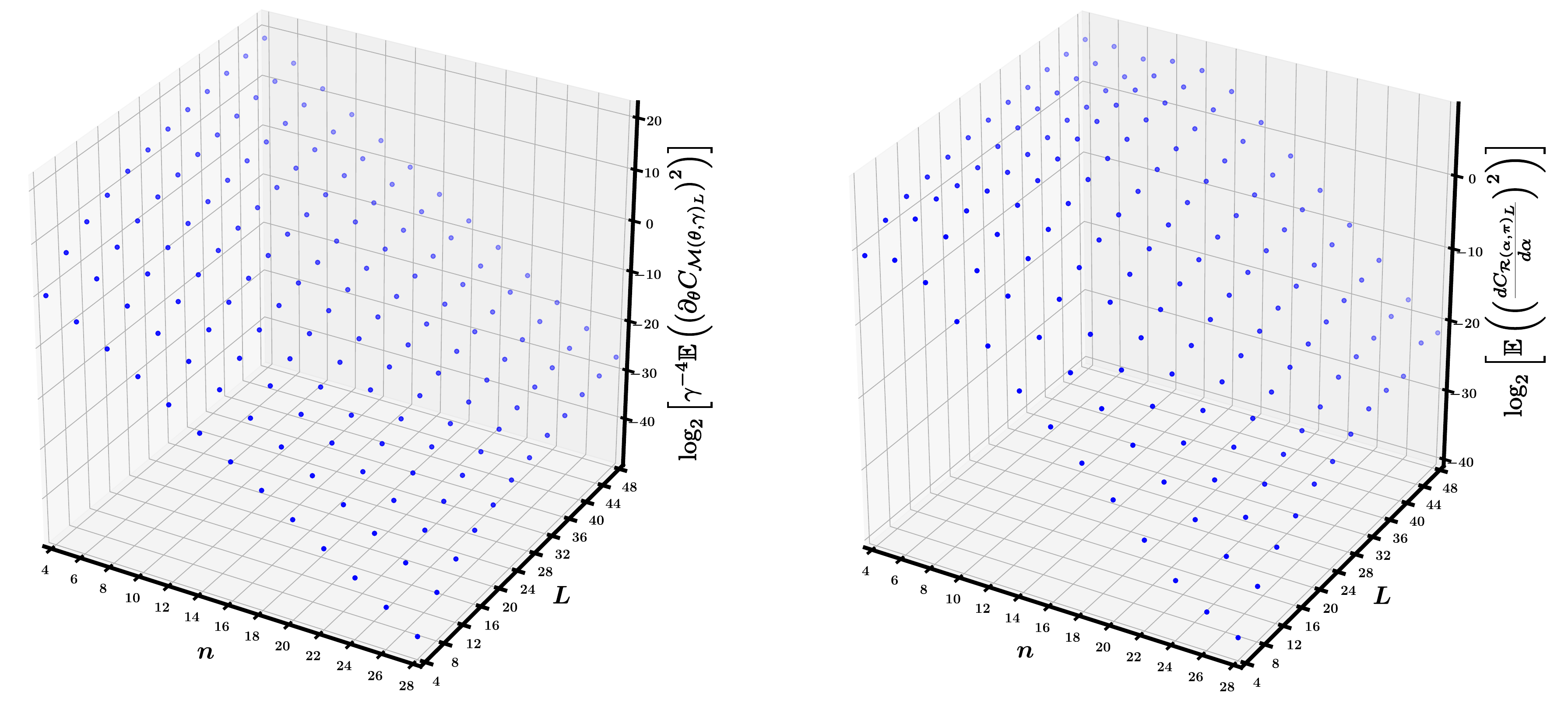}
    \caption{(Left) Logarithm of the result of Monte Carlo integration ($n=4,6,\ldots, 28$ and $L=4,8,\ldots,48$) of the $\OC(\gamma^{4})$ contribution to $E\left( \left( \del_{\theta}C_{\MC(\theta,\gamma)_{L}} \right)^{2}\right)$. Each data point is the mean of 20,000 samples from $[0,2\pi)$.  (Right) Logarithm of the result of numerical integration ($n=4,6,\ldots,24$ and $L=4,8,\ldots,48$) or Monte Carlo integration ($n=26,28$ and $L=4,8,\ldots,48$) of $E\left( \left( {dC_{\RC(\alpha,\pi)_{L}}\over d\alpha} \right)^{2}\right)$.}
\label{fig:tt}
\end{figure*}

\subsection{Less efficient version of variational Grover's algorithm\label{sec:badgrov}}

To gain analytical insight into the BPL phenomenon in the variational version of Grover's algorithm, we first analyze a less efficient version of Grover's algorithm first introduced in \cite{rieffel}. In this algorithm, an $n$-qubit register is first prepared in a superposition $\ket{\phi_{1}}={1\over \sqrt{2}}(\ket{m_{y}=-n/2}+\ket{m_{y}=n/2})$ of the lowest and highest weight eigenvectors of $J_{y}$ (in the $n$-qubit, i.e., spin-$n/2$, representation of $SU(2)$). This initial state is acted upon by layers of the unitary $S=e^{-i{2\pi\over n}J_{y}}e^{i\gamma C}e^{-i{2\pi\over n}J_{y}}e^{-i\gamma C}$, where $\gamma \in (0,\pi)$ and $C:=\dya{0}^{\otimes n}$ is the generator of the Grover oracle, i.e., the solution bitstring. Instead of quantifying the algorithm performance by using the distance of $S^{L}\ket{\phi_{1}}$ to the solution bitstring $\ket{0}^{\otimes n}$, the algorithm performance is quantified by the distance of $S^{L}\ket{\phi_{1}}$ to $\ket{\phi_{2}}$, where $\ket{\phi_{2}}$ is the zero eigenvector of $J_{y}$. There are two reasons for quantifying algorithm performance in this way: (1) $S^{n/2}$ is a unitary operator that, at $\OC(\gamma)$, acts as a rotation in the two-dimensional subspace spanned by $\ket{\phi_{1}}$, $\ket{\phi_{2}}$, (2) The squared modulus of the amplitude for the solution bitstring $\ket{0}^{\otimes n}$ in the state $\ket{\phi_{2}}$ is ${1\over 2^{n}}{n\choose n/2} \sim \left({2\over \pi n}\right)^{1/2}$, so if the algorithm produces the state $\ket{\phi_{2}}$, the solution bitstring can be obtained probabilistically with expected success rate $\OC(n^{-1/2})$ \cite{rieffel}. One can compare this success rate to the success rate $\OC(1)$ in Grover's original algorithm, and also compare the  $\OC(2^{{n\over 2}+{1\over 4}\log_{2}n - \log_{2}\gamma})$ required number of iterations to the $\OC(2^{n/2})$ iterations in Grover's original algorithm. The less efficient version of Grover's algorithm  described here is still an improvement over classical unstructured search.

We consider a variational version of this algorithm by taking the RPQC (Fig.~\ref{fig:first}c) \begin{equation}\MC(\theta,\gamma)_{L}=\displaystyle \overleftarrow{\prod_{j=1,\ldots,L/2}} e^{i\theta_{2j}J_{y}}e^{-i\gamma C}e^{i\theta_{2j-1}J_{y}}e^{i\gamma C}
\label{eqn:varbg}
\end{equation} ($L\equiv 0\;\text{mod}\;4$ so that an even number of iterations are used), and keeping the same initial state $\ket{\phi_{1}}$ and desired final state $\ket{\phi_{2}}$. The cost function is a function of $\theta=(\theta_{1},\ldots , \theta_{L})$, $\gamma$ and $L$, and is given to $\OC(\gamma^{2})$ by
\begin{eqnarray}
C_{\MC(\theta,\gamma)_{L}}&:=&1-\vert \langle \phi_{2} \vert \MC(\theta,\gamma)_{L}\vert \phi_{1}\rangle\vert^{2} \nonumber \\
&=&1-\gamma^{2}{ {n\choose n/2}\over 2^{2n-1}}\Big\vert \sum_{\ell=0}^{L-1}(-1)^{\ell}\cos {n\tilde{\theta}_{\ell}\over 2}   \Big\vert^{2} + \OC(\gamma^{4})
\label{eqn:bgcost}
\end{eqnarray}
where $\tilde{\theta_{\ell}}=\sum_{k=1}^{\ell}\theta_{k}$ are the partial sums of the variational parameters, and $\tilde{\theta}_{0}:= 0$ (see \ref{sec:bgcostproof} for proof of (\ref{eqn:bgcost})). Note that the cost function $C_{\MC(\theta,\gamma)_{L}}$ has the form in (\ref{eqn:cf1}) if one takes $O=\mathbf{I}-\dya{\phi_{2}}$, which is a global projection. If the parameters $\theta_{j}$ are perfectly correlated to a single varying parameter $\theta$, as they are in in Ref.~\cite{rieffel}, the expression in (\ref{eqn:bgcost}) simplifies to
\begin{equation}
C_{\MC(\theta,\gamma)_{L}}=1-\gamma^{2}{ {n\choose n/2}\over 2^{2n-1}} {\sin^{2}{(L-1)n\theta \over 4}\sin ^{2}{Ln\theta \over 4}\over \cos^{2}{n\theta \over 4}} .
\label{eqn:cfeval}
\end{equation}  
From (\ref{eqn:cfeval}), it follows that the cost function is minimized at $\OC(\gamma^{2})$ for $\theta = {2\pi \over n}$, which is the value used in Ref.~\cite{rieffel}. The cost function in (\ref{eqn:cfeval}) further satisfies $E\left( {dC_{\MC(\theta,\gamma)_{L}}\over d\theta} \right)= 0$ with respect to the uniform measure on $[0,2\pi)$. In \ref{sec:bgcostproof}, we also show that
\begin{equation}
\left( \del_{\theta}C_{\MC(\theta,\gamma)_{L}}\right)^{2} \le {{n\choose n/2}^{2}n^{2}\gamma^{4}\over 2^{4n-2}}{4L^{6}\over 9}.
\label{eqn:ffff}
\end{equation} Using (\ref{eqn:ffff}) in (\ref{eqn:cheby}) gives
\begin{equation}
P\left( \Big\vert \del_{\theta}C_{\MC(\theta,\gamma)_{L}} \Big\vert \ge\epsilon \right)\le {4{n\choose n/2}^{2}n^{2}L^{6}\gamma^{4}\over 9\epsilon^{2}2^{4n-2}}
\label{eqn:bgfin}
\end{equation}
with respect to the uniform measure on $[0,2\pi)$. From the asymptotic equality ${n \choose n/2} \sim {2^{n}\over \sqrt{\pi n/2}}$, one concludes that if $L\sim 2^{cn - \log_{2}n}$ and $0<c<1/3$, then the right hand side of (\ref{eqn:bgfin}) is $\OC(b^{-n})$ with $b=(1/4)^{1-3c}$. This fact  implies that at $\OC(\gamma^{2})$, the cost function (\ref{eqn:cfeval}) exhibits the BPL phenomenon if the RPQC is not taken to have sufficient depth.

To determine a critical RPQC depth for which the BPL phenomenon is absent, we note that if $L$ is such that $E\left( \Big\vert \del_{\theta}C_{\MC(\theta,\gamma)_{L}}\Big\vert\right)$ is bounded away from zero as $n\rightarrow \infty$, then $C_{\MC(\theta,\gamma)_{L}}$ does not exhibit the BPL phenomenon. In the present case, the fact that ${dC_{\MC(\theta,\gamma)_{L}}\over d\theta}$ is positive on the interval $(2\pi{(L-2)\over n(L-1)},{2\pi\over n}) $ implies the following lower bound

\begin{eqnarray}
E\left( \Big\vert \del_{\theta}C_{\MC(\theta,\gamma)_{L}} \Big\vert  \right)&\ge& {1\over 2\pi}\int_{2\pi{(L-2)\over n(L-1)}}^{2\pi/n}d\theta \, \del_{\theta}C_{\MC(\theta,\gamma)_{L}} \nonumber \\
&=& {{n\choose n/2}\gamma^{2}nL^{2}\over \pi 2^{2n}}\nonumber \\
&\sim& {\sqrt{2n}L^{2}\over \pi^{3/2}2^{n}}.
\label{eqn:bgfin2}
\end{eqnarray}
The asymptotic inequality (\ref{eqn:bgfin2}) shows that the expectation is bounded away from zero if $L\sim 2^{cn - {1\over 2}\log_{2}n}$, with $c\ge 1/2$. One concludes that the circuit $\MC(\theta,\gamma)_{L}$ can be efficiently optimized by gradient descent methods at circuit depths comparable to those required for successful algorithm performance. In Fig. \ref{fig:tt}, the integral $\gamma^{-4}E\left( \left(\del_{\theta}C_{\MC(\theta,\gamma)_{L}}\right)^{2} \right)$ is computed by Monte Carlo integration. The $\OC(L^{a}2^{-bn})$ scaling behavior is observed, which supports our prediction of an asymptotic crossover from BPL phenomenon at low $L$ to absence of BPL phenomenon at high $L$. 

Returning to the more general cost function (\ref{eqn:bgcost}), the question remains whether it is possible to avoid the BPL phenomenon when the angles $\theta_{k}$ characterizing the rotation layers $\lbrace e^{i\theta_{k}J_{y}} \rbrace_{k=1,\ldots ,L-1}$ vary independently. To show that it is not possible to avoid the BPL phenomenon for (\ref{eqn:bgcost}), consider a circuit depth $L \sim (2^{n/2 - {1\over 2}\log_{2}n})$ which we have shown allows to avoid BPL if the angles $\theta_{k}$ are perfectly correlated. For a parameter $\theta_{k}$ situated at a layer $k$ in the circuit that satisfies $L-k \sim (2^{cn})$ with $0<c<1/2$, one finds that 
\begin{eqnarray}
\Big\vert \partial_{\theta_{k}}C_{\MC(\theta,\gamma)_{L}} \Big\vert^{2} &\le & {{n\choose n/2}^{2}\gamma^{4}n^{2}(L-1)^{2}(L-k)^{2}\over  2^{4n-2}} \nonumber \\
&\sim& {2\over \pi}\gamma^{4}\left( \left({1\over 2}\right)^{1-c}\right)^{n}
\end{eqnarray}

From (\ref{eqn:cheby}), it is then clear that $P(\Big\vert  \del_{\theta_{k}}C_{\MC(\theta,\gamma)_{L}}\Big\vert  \ge \epsilon)$ is exponentially vanishing with $n$, which implies the BPL phenomenon for such parameters. Therefore, allowing the angles in $\theta$ to be uncorrelated implies that an exponential number of them are untrainable.

\subsection{Variational Grover's algorithm}

In Ref.~\cite{rieffel}, it was shown that for initial state $\ket{\phi_{1}}$, the RPQC $\RC(\alpha,\gamma)_{L}$ in (\ref{eqn:ggrpqc}) with $\alpha_{k}=2\pi/n$ and $\gamma_{k}=\pi$ for all $k$ obtains asymptotic fidelity 1 with the solution bitstring $\ket{0}^{\otimes n}$ when $L\sim 2^{n/2}$, thereby achieving the optimal oracle complexity for quantum unstructured search. We now consider the presence of the BPL phenomenon for the case of perfectly temporally correlated local rotations $\alpha_{k}=\alpha\in [0,2\pi)$ for all $k$,  while maintaining the structure of the optimal oracle in Ref.~\cite{rieffel} ($\gamma_{k}=\pi$ for all $k$). The cost function is defined by
\begin{equation}
C_{\RC(\alpha,\pi)_{L}}=1-\vert  \bra{0}^{\otimes n}\RC(\alpha,\pi)_{L}\ket{\phi_{1}}\vert^{2}
\label{eqn:cfgg}
\end{equation}
as a function of $\alpha$, $L$, and $n$ only.

Figure~\ref{fig:tt} shows the logarithm of the second moment of ${dC_{\RC(\alpha,\pi)_{L}}\over d\alpha}$. For each $L$, the asymptotically linearly decreasing behavior with $n$ indicates that for a fixed number of layers, the second moment of the derivative of the cost function decays exponentially with $n$. Therefore, for a fixed depth, the cost function (\ref{eqn:cfgg}) exhibits the BPL phenomenon. The remaining task is to determine a functional dependence of the depth $L$ on problem size $n$ that avoids BPL for the cost function (\ref{eqn:cfgg}).

To do this, we note that for $n \ge 16$, a least squares fit of the second moment of the derivative of (\ref{eqn:cfgg}) to the function $bL^{a}$ produces $a\approx 5$. This result is also observed to hold for the cost function (\ref{eqn:cfeval})  for all $n$. On the other hand, for every $L$, a least squares fit of the second moment of the derivative of (\ref{eqn:cfgg}) to the function $d2^{-rn}$ for $n\ge 14$ produces $r\approx 1.8$, whereas the same fit for second moment of the derivative of  (\ref{eqn:cfeval}) gives $r\approx 1.9$. These results allow us to conclude that the second moments of the gradients of both (\ref{eqn:cfeval}) and (\ref{eqn:cfgg}) are well fit by a scaling function of the form $(\text{const.}){L^{5}\over 2^{1.8n}}$ for $n\ge 14$. It follows that the crossover from BPL behavior to trainability carries over from the less efficient variational Grover's algorithm defined by (\ref{eqn:varbg}) to the efficient variational Grover's algorithm defined by $\RC(\alpha,\pi)_{L}$. In particular, if $L\sim  2^{cn}$ with $c<1.8/5= 0.36$, the BPL phenomenon is encountered. For the less efficient version of variational Grover's algorithm, this critical value of $c=0.36$ derived from a fitting analysis can be compared to the analytical estimate of the critical value $c=1/3$  derived from (\ref{eqn:bgfin}).

\section{Conclusions}

Strategies for efficient optimization of variational quantum algorithms, and quantum neural networks in general, are a prerequisite for the success of machine learning methods in quantum computation. We have shown that correlation of RPQC parameters spatially or temporally can be used to mitigate or avoid barren plateau landscapes (BPL) in specific variational quantum algorithms. The main consequence of correlation of RPQC parameters is the reduction in volume (in quantum state space) that can be accessed by the RPQC. Therefore, for generic variational quantum algorithms with a given RPQC structure, a tradeoff is expected between algorithm complexity and trainability via parameter correlation. This tradeoff can be quantified by, e.g., the expressibility of the RPQC \cite{yamamoto,guzikex}. An example of this tradeoff is provided by our analysis of a global cost function for the quantum approximate optimization algorithm in \ref{sec:rod}, which requires exponential depth in order to avoid the BPL phenomenon for a correlated RPQC. However, for variational versions of quantum unstructured search (Section \ref{sec:uns}), which require exponential circuit depth even in the optimal case, the use of correlated RPQCs increases trainability while maintaining algorithm performance and complexity.

The present results broaden the set of available strategies for defining efficient variational quantum algorithms for near term quantum processors. In future work, it would be interesting to explore the idea of correlating parameters as a pre-training approach (followed by training where the correlation is relaxed) in variational quantum algorithms.

\section{Acknowledgements}

We thank Lukasz Cincio for useful discussions related to translation invariant quantum circuits. The authors acknowledge support from LANL's Laboratory Directed Research and Development (LDRD) program. PJC also acknowledges support from the LANL ASC Beyond Moore's Law project, and from the U.S. Department of Energy, Office of Science, Office of Advanced Scientific Computing Research, under the Accelerated Research in Quantum Computing (ARQC) program.

\section*{References}

\bibliographystyle{unsrt_style}
\bibliography{phasebib}

\pagebreak

\appendix

\section{\label{sec:app1}Proof of (\ref{eqn:cvv})}

Taking $\theta_{\nu}$ to be the parameter of interest, one finds from 
(\ref{eqn:glob}) that 
\begin{equation}
E\left(   \left(\del_{\nu}C\right)^{2} \right) = n^{2}\int d\mu(\theta)\cos^{2(2n-1)}\left( {\sum_{i=1}^{L}\theta_{i}  \over 2} \right) \sin^{2}\left( {\sum_{i=1}^{L}\theta_{i} \over 2} \right)
\end{equation}
where $d\mu(\theta)={\prod_{i=1}^{L}d\theta_{i}\over (2\pi)^{L}}$. Making the linear change of variables $v_{j}=\sum_{i=1}^{j}\theta_{i} \;\text{mod}\; 2\pi$, $j=1,\ldots ,L$, which changes neither the measure nor the domain of integration, one obtains the integral
\begin{eqnarray}
E\left(   \left(\del_{\nu}C\right)^{2} \right) &=&{n^{2}}\int {dv_{L}\over 2\pi}\cos^{2(2n-1)}\left( {v_{L}  \over 2} \right) \sin^{2}\left( {v_{L} \over 2} \right) \nonumber \\
&=& {2n^{2} \over 4n-1}\int_{-{\pi \over 2}}^{\pi \over 2}{du \over 2\pi}\cos^{4n}u \nonumber \\
&=& {n^{2} \over 2^{4n}(4n-1)}{4n\choose 2n} \nonumber \\
&\sim& {n^{1\over 2}\over 4\sqrt{2\pi }}
\end{eqnarray}
where $u={v_{n}\over 2}$ and we use the asymptotic form ${2\ell \choose \ell} \sim {2^{2\ell}\over \sqrt{\pi \ell}}$ for the central binomial coefficient. Note that the result is independent of $L$.

\section{\label{sec:bgcostproof}Proof of (\ref{eqn:bgcost}) and (\ref{eqn:ffff})}

We define the partial sums $\tilde{\theta}_{j}:=\sum_{k=1}^{j}\theta_{k}$, $j=1,\ldots,L$ (taking $\tilde{\theta}_{0}=0$ by definition), and the $SU(2)$ coherent states $\ket{z}:=\left( 1+\vert z\vert^{2}\right)^{-n/2}e^{zJ_{-}}\ket{0}^{\otimes n}$, where $J_{-}:=\sum_{j=1}^{n}\sigma_{-}^{(j)}$. We make use of the following rotation formula that holds for $a\in \mathbf{R}$:
\begin{eqnarray}
e^{2iaJ_{y}}\ket{z}=\Big \vert {z\cos  \vert a\vert -{a\sin  \vert a\vert\over \vert a\vert} \over z{a\sin  \vert a\vert\over \vert a\vert}+\cos  \vert a\vert}\Big \rangle 
\label{eqn:aaa}
\end{eqnarray}
and the SU(2) coherent state inner product in the spin $j=n/2$ representation
\begin{equation}
\langle z\vert z'\rangle = \left( (1+\vert z\vert^{2})(1+\vert z'\vert^{2})\right)^{-n/2}\left( 1+\overline{z}z'\right)^{n}.
\label{eqn:bbb}
\end{equation}
For small $\gamma$, the random parameterized quantum circuit for $\MC(\theta,\gamma)_{L}$ appearing in (\ref{eqn:varbg}) can be written
\begin{eqnarray}
\MC(\theta,\gamma)_{L}&=&e^{-i\theta_{L}J_{y}}\left( 1+i\gamma\ket{z=0}\bra{z=0}\right)\nonumber \\ &{} &e^{-i\theta_{L-1}J_{y}}\left( 1-i\gamma\ket{z=0}\bra{z=0}\right) \nonumber \\
&{}& \vdots \nonumber \\
&{}& e^{-i\theta_{2}J_{y}}\left( 1+i\gamma\ket{z=0}\bra{z=0}\right)\nonumber \\ &{}& e^{-i\theta_{1}J_{y}}\left( 1-i\gamma\ket{z=0}\bra{z=0}\right).
\end{eqnarray}
At $\OC(\gamma^{2})$, the cost function is given by
\begin{eqnarray}
C_{\MC(\theta,\gamma)_{L}}&=&1-{\gamma^{2}\over 2^{n}}{n\choose n/2}\big\vert -\langle z=0 \vert \phi_{1} \rangle + \langle z=0 \vert e^{-i\tilde{\theta}_{1}J_{y}}\vert \phi_{1} \rangle \nonumber \\
&-& \langle z=0 \vert e^{-i\tilde{\theta}_{2}J_{y}}\vert \phi_{1} \rangle +\cdots - \langle z=0 \vert e^{-i\tilde{\theta}_{L-1}J_{y}}\vert \phi_{1} \rangle \big\vert^{2}
\label{eqn:ccc}
\end{eqnarray}
where we have used the fact that $\left( \ket{\phi_{2}},e^{i\varphi J_{y}} \ket{z=0} \right) = {1\over 2^{n/2}}\sqrt{{n\choose n/2}}$ for any $\varphi \in [0,2\pi)$. We now use (\ref{eqn:aaa}) and (\ref{eqn:bbb}) to evaluate
\begin{eqnarray}
\langle z=0 \vert e^{-i\varphi J_{y}}\vert b_{+}\rangle &=& \left(  {\left(1-i\tan {\varphi\over 2 }\right)^{n} \over \left( 1+ \tan^{2} {\varphi\over 2 }\right) 2^{n\over 2}} +{\left( 1+i\tan {\varphi\over 2 }\right)^{n} \over \left( 1+ \tan^{2} {\varphi\over 2 }\right) 2^{n\over 2}}  \right) \nonumber \\
&=& {\sqrt{2}\over 2^{n/2}}\cos {n\varphi \over 2}.
\label{eqn:ip2}
\end{eqnarray}
Using this formula in (\ref{eqn:ccc}), one obtains (\ref{eqn:bgcost}). 

Equation~(\ref{eqn:ffff}) is proved by writing 
\begin{eqnarray}
\Bigg\vert \del_{\theta}C_{\MC(\theta,\gamma)_{L}} \Bigg\vert &\le & {{n\choose n/2}\gamma^{2}\over 2^{2n-1}}\left[ \Bigg\vert {(L-1)n\sin {(L-1)n\theta \over 4} \cos {(L-1)n\theta \over 4}\sin^{2} {Ln\theta \over 4} \over 2\cos^{2}{n\theta \over 4} }\Bigg\vert \right. \nonumber \\
&{}&  \left. +\Bigg\vert {Ln\sin^{2} {(L-1)n\theta \over 4} \cos {Ln\theta \over 4}\sin {Ln\theta \over 4} \over 2\cos^{2}{n\theta \over 4} }  \right. \nonumber \\
&{}& \left.  +{ n \sin^{2} {(L-1)n\theta \over 4} \sin^{2} {Ln\theta \over 4}\sin {n\theta \over 4}  \over 2\cos^{3}{n\theta \over 4}} \Bigg\vert \right] \nonumber \\
&\le&  {{n\choose n/2}\gamma^{2}\over 2^{2n-1}}\left[ {L^{2}(L-1)n\over 2}+\Big\vert g(\theta)\Big\vert \right] \nonumber \\
&\le & {{n\choose n/2}\gamma^{2}n\over 2^{2n-1}}{2L^{3}\over 3} 
\end{eqnarray}
where 
\begin{equation}
g(\theta):={Ln \cos {Ln\theta \over 4}\cos{n\theta \over 4} +  n  \sin {Ln\theta \over 4}\sin {n\theta \over 4} \over 2\cos^{3}{n\theta \over 4} }  
\end{equation}
and where in the first and second inequalities we used $\text{max}_{\theta \in [0,2\pi)}\Big\vert {\sin mn\theta/4 \over \cos n\theta/4 }\Big\vert = m$ for $m\equiv 0\;\text{mod}\;4$. To derive the third inequality, we note that $\theta = {2\pi\over n}$ is a critical point of $g(\theta)$ and it is the only zero of the denominator of $g(\theta)$ in the period $4\pi \over n$. Like the Dirichlet kernel, the global maximum of $g(\theta)$ is expected to occur at a zero of the denominator. One finds that $\vert g({2\pi\over n})\vert = {Ln(L^{2}-1)\over 6}$.

\section{\label{sec:rod}Avoiding BPL in the ring of disagrees algorithm}

The RPQC has the quantum alternating operator ansatz (QAOA) form consisting of alternating driver and mixer layers
\begin{equation}
\MC(\beta,\gamma)_{L}:=\displaystyle\overleftarrow{\prod_{k=1,\ldots,L}}e^{-i\beta_{k}J_{x}}e^{-i\gamma_{k}C}
\label{eqn:mixdriv}
\end{equation} with $\gamma_{j}\in [0,\pi/2 )$ and $\beta_{j}\in [-\pi,\pi)$ and $C={n\over 2}-{1\over 2}\sum_{j=1}^{n}\sigma_{z}^{(j)}\otimes \sigma_{z}^{(j+1)}$ (see Fig.~\ref{fig:first}d). A 2-local cost function is defined by 
$C^{(L)}(\beta,\gamma)=\langle \psi(\beta ,\gamma) \vert C \vert \psi(\beta ,\gamma) \rangle$ where $\ket{\psi(\beta, \gamma)}=\MC(\beta ,\gamma)_{L}\ket{\psi_{0}}$ and $\ket{\psi_{0}}$ is the maximal eigenvector of $J_{x}$ on $n$ qubits (i.e., $\propto (\ket{0}+\ket{1})^{\otimes n}$). In the ring of disagrees problem, one seeks parameters in (\ref{eqn:mixdriv}) such that $C^{(L)}$ is maximized. Here we consider the BPL phenomenon for the mixer parameters $\beta_{j}$ for small values of the driver parameters $\gamma_{j}$. The contribution to $C^{(L)}$ of linear order in the $\gamma_{j}$ is given by 
\begin{equation}
C^{(L)}= {n\over 2}+ {n\over 2}\sum_{j=0}^{L-1}\gamma_{j}\sin 2x_{j} +\sum_{i\le j}\OC(\gamma_{i}\gamma_{j})
\end{equation}
where $x_{j}=\beta_{j+1}+\cdots + \beta_{L}$. It is evident that there is no BPL for $C^{(L)}$ due to the fact that $E\left( \vert \del_{\beta_{j}}C^{(L)}\vert \right) \in \OC(n)$ for all $j$. Therefore, $C^{(L)}$ allows to efficiently train the mixer angles $\beta_{j}$ in the ring of disagrees problem when the driver layers consist of short time evolutions generated by $C$. 

For comparison, a global cost function that again identifies the maximal ring of disagrees is given by taking $C^{(G)}=\langle \psi(\beta ,\gamma) \vert O \vert \psi(\beta ,\gamma) \rangle$ with $O=\sum_{k=1}^{2}\ket{\psi_{k}}\bra{\psi_{k}}$, where $\ket{\psi_{1}}=\left( \ket{0}\otimes \ket{1}\right)^{\otimes n/2}$, $\ket{\psi_{2}}=\left( \ket{1}\otimes \ket{0}\right)^{\otimes n/2}$, i.e., the degenerate highest eigenvectors of $C$.  Using uncorrelated mixer angles $\beta_{j}$ implies that a subset of mixer angles exhibits BPL regardless of circuit depth (cf. the analogous observation in Section \ref{sec:uns}). We therefore correlate the driver parameters (i.e., $\gamma_{j}=\gamma$ for all $j$) and the mixer parameters (i.e., $\beta_{j}=\beta$ for all $j$), and expand to linear order in $\gamma$. The result is given by
\begin{equation}
C^{(G)}={1\over 2^{n-1}}-{n\gamma \over 2^{n}}\left( {\cos((2L+1)\beta) - \cos \beta \over \sin \beta}\right) + \OC(\gamma^{2})\,,
\end{equation}
for which one finds that
\begin{eqnarray}
E\left( \left( \del_{\beta}C^{(G)}\right)^{2}\right)&\sim& {(2L+1)^{2}n^{2}\gamma^{2}\over 2^{2n}}\int_{-\pi}^{\pi}{d\beta \over 2\pi}{\sin^{2}((2L+1)\beta)\over \sin^{2}\beta} \nonumber \\
&=& {(2L+1)^{3}n^{2}\gamma^{2}\over 2^{2n}}
\label{eqn:llimqaoa}
\end{eqnarray}
where we have used the $L\rightarrow \infty$ asymptotic in the first line, and the $L^{1}$ norm of the Fej\'{e}r kernel in the second line of (\ref{eqn:llimqaoa}). Therefore, BPL is encountered if $L\sim 2^{cn-{2\over 3}\log_{2}n}$ with $c<2/3$.  Conversely, the logarithmic divergence of the $L^{1}$ norm of the Dirichlet kernel implies that
\begin{eqnarray}
E\left( \Big\vert \del_{\beta}C^{(G)}\Big\vert \right) &\sim &{(2L+1)n\gamma \over 2^{n}}\int_{-\pi}^{\pi}{d\beta \over 2\pi}\Big\vert {\sin (2L+1)\beta \over \sin \beta} \Big\vert \nonumber \\
&\ge& {(2L+1)n\gamma \ln (2L+1) \over \pi^{2}2^{n-1}},
\end{eqnarray}
from which it follows that no BPL is encountered if $L\ln L$ scales as $2^{n-\log_{2}n}$.

Note that a proof of computational universality of QAOA \cite{lloydopt} requires that the $(\beta_{j},\gamma_{j})$ can be varied, so the question remains whether the quantum approximate optimization approach to MaxCut still works for an RPQC defined by QAOA having correlated driver or mixer layers. 
For the ring of disagrees problem with $4\le n \le 10$, we find numerically that $\max_{\beta,\gamma}C^{(L)}(\beta ,\gamma)$ with ($\beta_{1}=\ldots =\beta_{L}$ and $\gamma_{1}=\ldots = \gamma_{L}$) is increasing with $L$, suggesting that in this case, layer-correlated QAOA provides a suboptimal, but low-dimensional, approach to the ring of disagrees problem.

\end{document}